# Ultrafast electronic and lattice dynamics in laser-excited crystalline bismuth

Alexey A. Melnikov[1], Oleg V. Misochko[2], Sergey V. Chekalin[1]

[1]*Institute of Spectroscopy Russian Academy of Sciences, Troitsk, Moscow Oblast, 142190, Russia*
[2]*Institute of Solid State Physics Russian Academy of Sciences, Chernogolovka, Moscow Oblast, 142432, Russia*

Email: melnikov@isan.troitsk.ru

## Abstract

Femtosecond spectroscopy is applied to study transient electronic and lattice processes in bismuth. Components with relaxation times of 1 ps, 7 ps and ~ 1 ns are detected in the photoinduced reflectivity response of the crystal. To facilitate the assignment of the observed relaxation to the decay of particular excited electronic states we use pump pulses with central wavelengths ranging from 400 nm to 2.3 μm. Additionally, we examine the variation of parameters of coherent $A_{1g}$ phonons upon the change of excitation and probing conditions. Data analysis reveals a significant wavevector dependence of electron-hole and electron-phonon coupling strength along $\Gamma$-$\Lambda$-$T$ direction of the Brillouin zone.

## 1. Introduction

Crystalline bismuth has been studied by methods of ultrafast optics for already two decades. The focus of the research is mainly on structural phenomena associated with coherent atomic oscillations [1–5] and phase transitions initiated by femtosecond laser pulses [6–8]. The longstanding interest to ultrafast effects in bismuth is due to the unique properties of the crystal [9–11]. Bismuth is characterized by rhombohedral A7 lattice, which is very close to simple cubic one: the relative difference of main lattice parameters – trigonal shear angle and internal displacement – is of the order of 5% for these two structures [11]. As a result, bismuth is exceptionally sensitive to external perturbations tending to approach the phase of a higher symmetry. In particular, it is possible to induce atomic displacements of about 5-8% of the nearest neighbor distance exposing bismuth to femtosecond laser pulses with intensity approaching melting threshold [6]. The highly excited vibrational states are often used to search for various unconventional and nonclassical lattice effects like subpicosecond disordering [8], collapse and revival of coherent oscillations [12, 13], transient Bose-Einstein condensation of phonons [14], phase rigidity [15] and squeezed phonon states [16–18].

As compared to the lattice dynamics and in contrast to other elemental solids ultrafast electronic dynamics in bismuth have received considerably less attention. In a few works the characteristic relaxation times of photoexcited charge carriers were obtained, but both the reported values and their interpretation varied considerably. However, the information on transient electronic processes is important. First of all, it is indispensable for the correct interpretation of ultrafast structural effects in bismuth as in a Peierls distorted solid. A proper description of electronic relaxation on the microscopic level will allow avoiding the situations when essentially quantum phenomena are treated phenomenologically, or just with a reference to thermodynamical considerations [19–21].

Additionally, the study of electronic dynamics in crystalline bismuth is of great interest due to unique properties of its charge carriers. Electrons and holes in bismuth differ significantly from those in ordinary metals and semiconductors: the crystal is characterized by extremely small Fermi surface, only one free electron per $10^5$ atoms (~ $10^{18}$ cm$^{-3}$) with long mean free path (up to ~ 1 mm) and highly anisotropic effective mass [9, 10]. For the dilute gas of charge carriers with such specific features electronic interactions and correlations become important. Indeed, the recent studies of electrical and thermal transport claimed the observation of extraordinary behavior of Dirac electrons in bismuth and also atypical corrections to the Fermi liquid picture, such as electronic fractionalization and unusual collective modes [22–26]. Methods of ultrafast optics are in some ways complementary to transport measurements and can provide additional information on the lifetime of various elementary excitations as well as on the interactions between them.

The present work is an attempt to provide a better understanding of electronic and lattice dynamics in bismuth on a picosecond timescale applying broadband femtosecond pump-probe method in the visible and near infrared ranges. Concentrating on the photoinduced reflectivity response at multiple probe wavelengths we have managed to observe multicomponent electronic dynamics correlated with both coherent and non-coherent lattice dynamics spanning a wide range from pico- to nanoseconds. The analysis of pump wavelength dependence of the electronic decay has revealed significant variation of electron-hole and electron-phonon coupling strength along the $\Gamma$-$T$ direction of the Brillouin zone, demonstrating a singularity inherent for Peierls-distorted systems.

## 2. Methods

Femtosecond pump-probe technique is a powerful tool for the investigation of ultrafast processes in condensed matter. The method is based on time-resolved detection of short-lived changes of optical parameters (e.g. reflectivity or transmittance) induced in a sample by ultrashort laser pulses and is intended to define how fast is the relaxation of the transient excited state. Provided there are enough input parameters and a reliable model at hand, the observed decay rates may be related to lifetimes of elementary excitations and reveal the strength of interaction between quasiparticles – a key value for the solid state theory.

In order to study ultrafast processes in bismuth researchers frequently use the so-called degenerate layout, where both pump and probe pulses have the same central wavelength. In this case the response of a crystal to laser excitation is represented by only single spectral point, and thus unequivocal derivation of the electronic system state becomes rather difficult.

To clarify the dynamics one may also utilize X-ray or electronic diffraction [6,7, 27–30]; however, the corresponding methods provide almost no information on the dynamics of outer (valence) electrons. Recently to address the kinetics of individual Bloch states in bismuth scientists used ultrafast angle resolved photoelectron spectroscopy; yet, this approach deals mainly with surface states [31].

A simple way to obtain additional information on the ultrafast dynamics and to avoid ambiguous interpretation is to collect time-resolved data at multiple wavelengths. The shift to a broader spectral domain is usually performed with the help of nonlinear optical methods like parametric amplification, or white light continuum generation. Thereby (with the use of proper photodetectors) one gets an easy access to ultraviolet, visible, near- and mid-infrared ranges and the pump-probe method is then referred to as ultrafast spectroscopy.

In our experiments we used femtosecond pulses at 800 nm delivered by a Ti:Sapphire laser. A beam from the output was divided into two parts. The first part was sent either in a BBO crystal to provide pump pulses at 400 nm, or in a parametric amplifier to generate pulses covering wavelengths from 600 to 2500 nm. Duration of pump pulses in each case was about 70 fs. The second part was attenuated and used to generate femtosecond supercontinuum or acted as a probe pulse. The pump and the probe pulses were focused onto a monocrystalline bismuth sample almost perpendicular to its surface (cleaved perpendicular to (111) direction). The pump beam was modulated by an optical chopper and the energy of the reflected probe pulse was measured. Normalization to the energy of a reference pulse, that was reflected from an unperturbed part of the sample, gave the relative change of reflectivity $\Delta R/R(t)$ at a given delay time $t$ between pump and probe pulses. We had two measurement setups at our disposal. The first one (multi-channel) operated with femtosecond continuum as a probe pulse and pump pulses at 400 nm. In this case the wavelength resolved $\Delta R/R(\lambda, t)$ data were obtained with a temporal resolution of about 250 fs (the value of the pump and probe cross-correlation). The second setup (single-channel) used a set of fixed probe wavelengths granting a temporal resolution of about 100 fs. All measurements were made at atmospheric pressure and room temperature.

## 3. Results

Main features of the photoinduced response of bismuth measured in our experiments are illustrated by Fig. 1 and Fig. 2. The crystal was excited by femtosecond laser pulses at 400 nm with energy density of 1.8 mJ/cm$^2$, while the probe wavelength varied in the 450-780 nm interval. Contour plot on Fig. 1 corresponds to the two-dimensional array $\Delta R/R(\lambda, t)$ recorded by means of the multi-channel pump-probe setup. At every wavelength a typical form of the photoinduced response is well reproduced. Near zero time delay reflectivity sharply increases indicating the arrival of a pump pulse, while subsequent relaxation consists of a combination of a monotonic signal with damped oscillations superimposed. The oscillations are caused by collective fully symmetric atomic motion or, in other words, $A_{1g}$ coherent phonons. The coherent lattice vibrations modulate reflectivity of bismuth in all visible spectrum, while the amplitude of modulation varies across the 450-780 nm range being nonzero at every probe wavelength used. The monotonic component reflects the relaxation of photoexcited charge carriers and noncoherent (thermal) lattice processes.

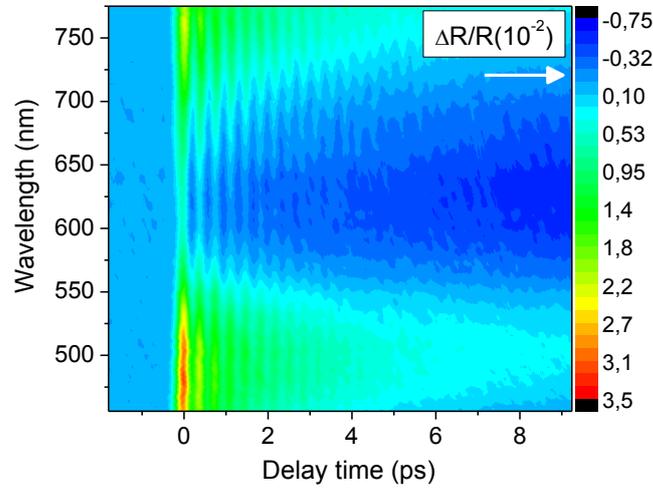

Fig. 1. Two dimensional array of *ΔR/R(λ, t)* measured using the excitation laser pulses at 400 nm with the fluence of 1.8 mJ/cm$^2$.

To better illustrate the photoinduced spectral changes, the decay traces measured at 600 and 780 nm probe wavelengths are plotted in Fig. 2. These data were recorded using single-channel pump-probe setup which has a higher temporal resolution and a better signal-to-noise ratio than the multi-channel system. In that way, the main features of ultrafast relaxation are made more distinct.

To separate monotonic part of the photoinduced response we used Fourier filtering – the corresponding traces are shown in Fig. 2 for two probe wavelengths: 600 and 780 nm. After the analysis of the whole range of probe wavelengths it was found that the simplest form of monotonic component is observed near the 600 nm probe wavelength. Upon ultrafast laser excitation the reflectivity of bismuth at 600 nm gradually decreases with characteristic time $\tau_2$, reaching a quasiconstant value at ~ 25 ps. With our experimental setups we were not able to measure nanosecond dynamics, however, the comparison with the results of previous studies at 800 nm shows that return to an equilibrium lasts roughly 1 ns [20]. Note, that such a comparison is reasonable because the long-term relaxation is reproduced at every probe wavelength in the whole visible range.

Indeed, it turns out that the trailing parts of all decay traces can be superposed by a parallel translation along vertical axis. The difference is observed only at the delay times ranging from 0 to ~ 4 ps and is due to the admixture of a faster component with characteristic relaxation time of $\tau_1$ (see the dashed line at the 780 nm decay trace in Fig. 2). A maximum contribution of this component is observed on the edges of the probed region, which cause two pronounced maxima near zero time delay on the contour plot in Fig. 1.

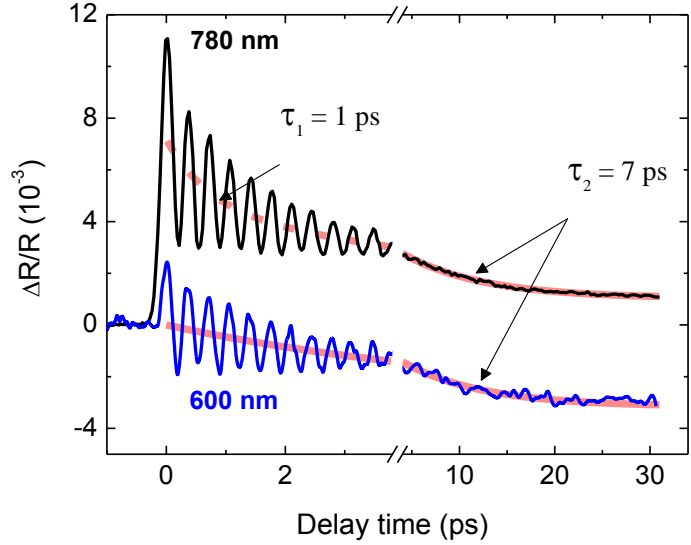

Fig. 2. Decay traces at 600 and 780 nm measured using single-channel detection and excitation at 400 nm wavelength with ~ 1.3 mJ/cm² fluence. Solid line corresponds to the slower component of the response while the dashed – to the faster.

According to the results presented above, we modeled monotonic part of the photoinduced response by

$$(\Delta R/R)_{mon} = \sigma_1 e^{-t/\tau_1} + \sigma_2 e^{-t/\tau_2} + \sigma_3, \quad (1)$$

where $\sigma_1$, $\sigma_2$ and $\sigma_3$ denote contributions of temporal components at a given probe wavelength. Note that the reliable measurement of multicomponent electronic decay is possible only due to the use of broadband probing. Indeed, it turns out that if the probe wavelength is near 600 nm (as in several early works on bismuth), the fastest picosecond component can hardly be detected since its contribution is close to zero.

Returning to Eq. 1, $\sigma_3$ must be, strictly speaking, multiplied by an exponent, but because its decay time is large (~ 1 ns), then, even in the maximal time window of 30 ps used in our experiments, the constant approximation remains reasonable. Fitting the measured data to Eq. 1, we found that $\tau_1$ = 1.0±0.3 ps, $\tau_2$ = 7.0±0.5 ps, and $\sigma_2 \approx 0.007$.

Let us now discuss the oscillating part of the photoinduced response, whose spectrum is shown in Fig. 3(a). The corresponding spectral line is characterized by a central frequency of about 2.9 THz, which is close to the frequency of the $A_{1g}$ phonon mode of bismuth (2.93 THz, [3]). The small discrepancy in frequency is often related to the "softening" of interatomic bonds due to photoexcited electrons [3–5, 32–35]. Checking the spectral dependence of phase and lifetime (~2 ps) of reflectivity oscillations we found that both are independent of probe wavelength.

The observed frequency shift is actually not stationary [5, 32]. Its time dependence becomes evident if we compare the oscillatory component with damped cosine function with frequency of 2.93 THz as shown in Fig. 3(c). One can choose an initial phase of the model cosine in such a manner, that at time delay values larger than 2 ps the model cosine coincides with the observed oscillations, while closer to zero time delay a considerable deviation is observed. To extract time dependent instantaneous frequency of coherent phonons $\nu(t)$, we measured phase shift $\Delta\phi$ between the oscillatory component and the damped cosine at points where the latter curve changes sign. The result is shown in the Fig. 3(b).

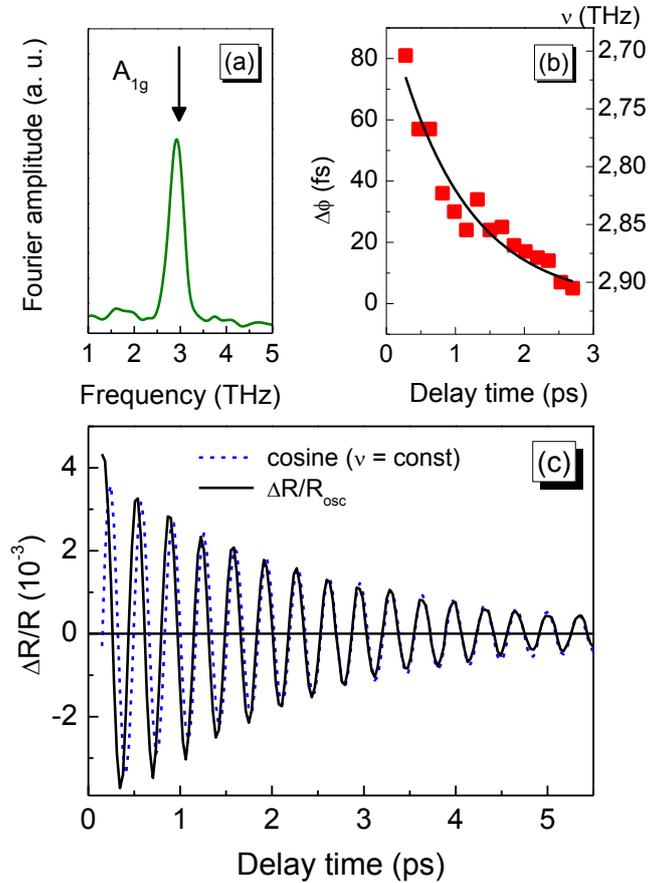

Fig. 3. (a) Fourier spectrum of oscillations in the photoinduced response. (b), (c) – Phase shift between oscillations (extracted from the decay trace at $\lambda_{probe} = 700$ nm) and a cosine of constant frequency. Squares show the calculated instantaneous frequency $\nu(t)$.

In this way we found that temporal dependence of $\Delta\phi$ can be satisfactorily fitted by the exponential decay with characteristic time $\tau = 1 \pm 0.3$ ps. Then the instantaneous frequency $\nu(t)$ can be calculated as a temporal derivative of $\Delta\phi$, and it also demonstrates exponential behavior. Immediately after the excitation, the frequency is 2.7 THz, gradually increasing to the unperturbed value of 2.93 THz. This temporal behavior naturally leads to a broadened spectrum with a shifted mean value. Taking into account these results the oscillating part of the photoinduced response is described by the following expression:

$$(\Delta R/R)_{osc} = Ae^{-\gamma t}\cos\left(\left(\nu_0 + \Delta\nu e^{-t/\tau}\right)t\right) \qquad (2)$$

where $A$ is the amplitude, $\gamma$ – the decay rate and $\nu_0$ – the unperturbed frequency of $A_{1g}$ oscillations.

It should be emphasized that the amplitude is the only quantity related to the oscillating part of the photoinduced response that depends on probe wavelength (the frequency and its chirp, as well as relaxation time are wavelength independent). The functional form of this dependence contains important information on the interaction between $A_{1g}$ vibrations and electrons. However, due to the fact that fully symmetric phonons couple to all electronic states in bismuth, a detailed theoretical analysis of this dependence is rather difficult and is beyond the scope of the present work.

In the present work we make use of the probe wavelength dependence $A(\lambda)$ to compare it with the spectral form of temporal components $\sigma_i(\lambda)$. This simple approach has something in common with various modulation spectroscopy methods, where external periodic perturbation (strain, electrical field, etc.) modifies electronic states and changes optical parameters. In the case of bismuth reflectivity spectrum is modulated by collective atomic motion initiated by an ultrafast laser pulse. Surprisingly, such a comparison helps to clarify the nature of electronic relaxation in bismuth.

One more important feature of the presented data is the similar character of the initial $\sigma_1$ electronic decay and the relaxation of the oscillation frequency shift $\Delta\nu$. This similarity is an evidence of strong coupling between electronic and lattice subsystems in bismuth, and as we will show in the next section, only a portion of all excited charge carriers occupying certain states of high symmetry take part in this interaction.

## 4. Discussion

*4.1. Single vs. multiple Fermi-Dirac distribution for charge carriers*

Photoinduced ultrafast processes in bismuth (and in many other crystalline solids) are often described by the so called two-temperature model [19–21, 28, 29, 36]. According to this model, femtosecond laser radiation is absorbed by a portion of valence electrons and a highly nonequilibrium electronic distribution is formed. Electron-electron scattering leads to the redistribution of energy among all charge carriers and a certain electronic temperature $T_e$ is established. This process is referred to as thermalization and usually takes from several to tens of femtoseconds to finish. Therefore, at the end of excitation pulse the crystal lattice is at room temperature $T_0$, while the electronic system has temperature of $T_e >> T_0$. The subsequent process is the energy transfer from hot charge carriers to the cold lattice involving electron-phonon interaction and it lasts much longer than the thermalization.

Yet, the two-temperature model implies monoexponential relaxation of both electronic and lattice subsystems and is therefore oversimplified. Indeed, previous experiments revealed two-exponential decay [37–39], while a three-component photoinduced response was observed in the present work. Nevertheless, the notion of hot gas of charge carriers is often used straightforwardly to interpret not only ultrafast electronic processes, but also collective structural rearrangements in bismuth. As lattice effects are concerned, the corresponding models originate from the concept of displacive excitation which describes the dynamics of $A_{1g}$ phonon mode (most affected by optical excitation). Taking into account the above, it is not surprising that the models generally operate by thermodynamic quantities and identify differently the force that drives coherent atomic motion (this force can originate from the population of higher lying states, the change of electronic temperature $T_e$ [19] or its nonzero gradient [20], or also the variation of entropy [21]).

Recently several attempts were made to obtain more detailed information on the principal ultrashort processes in bismuth. Here one should mention the papers in which discussion goes beyond a single Fermi-Dirac distribution for the electronic system. The corresponding models suggest that immediately after ultrafast excitation distinct thermal distributions are formed for the electrons in the conduction band and for the holes in the valence band [17, 27, 33, 39, 40]. A validity of such an approach is a matter of considerable debate, since several theoretical studies as well as X-ray measurements led to the negative conclusion as far as two chemical potential model is concerned [29, 41, 42]. Instead, it was proposed that characteristic thermalization time is the same for all charge carriers and is about 100 fs or even faster [29]. However, in a recent study scientists used THz probing to monitor excited electrons in bismuth close to the Fermi level and claimed that electron-hole equilibration is retarded due to large mass anisotropy and occurs in ~ 0.6 ps [39].

The results obtained in the present study by an alternative method of broadband femtosecond probing speak in favor of multicomponent electronic distribution. We treat the photoexcited state of the electronic system in bismuth as a composition of several groups of nonequilibrium charge carriers. Each group occupies particular states of the Brillouin zone close to the points of high symmetry and the properties of these states control corresponding relaxation rates. Somewhat similar approach has been presented in [39], but the suggested interpretation of electronic dynamics near the local band extrema, as well as the obtained intervalley relaxation time differs markedly from ours.

*4.2. Regimes of high and low excitation intensity*

The character of ultrafast electronic and lattice relaxation in photoexcited bismuth depend on the excitation regime as well as on the type of the sample (thin film or single crystal). This fact could not be ignored if one compares the results of measurements in order to arrive at some unified model.

First of all, as it is well known, the photoinduced response depends on the intensity of a pump laser radiation [43]. However, for bismuth the dependence is not sharp and therefore it is reasonable to speak about two limits of low and high excitation intensity. We used excitation fluence of about 1 mJ/cm$^2$, which corresponds to ~ $10^{20}$ cm$^{-3}$ of photoexcited charge density (note that the density of free charge carriers in bismuth is about $10^{18}$ cm$^{-3}$ at room temperature [9]). In this limit of high excitation intensity the higher fluence leads only to an increase of photoinduced reflectivity signal, while relative contributions of each relaxation component as well as the decay rates remain almost constant.

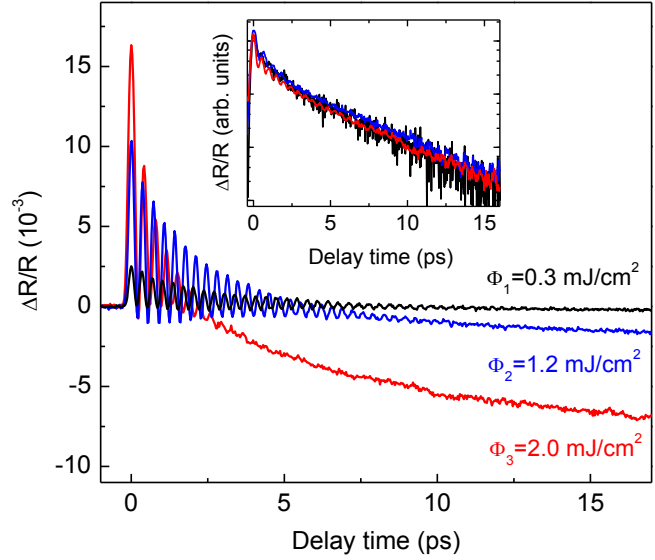

Fig. 4. Decay traces measured at three values of pump laser fluence. The inset shows the curves in a logarithmic scale with oscillations filtered out (prior to normalization traces were shifted to equalize their quasiconstant $\sigma_3$ components).

We measured photoinduced reflectivity response of bismuth at 700 nm probe wavelength using three different values of excitation laser fluence $\Phi$: 0.3, 1.2 and 2.0 mJ/cm$^2$ (pump pulses at 400 nm). Though $\Phi$ spans almost one order of magnitude, no considerable change of relaxational features is observed: both the fast ($\tau_1$) and intermediate ($\tau_2$) components have proportional contributions (see Fig. 4). Therefore, we can conclude that the decay processes with electronic density-dependent rate (e.g. Auger recombination) are not effective in bismuth at the high excitation level, at least on the picosecond timescale.

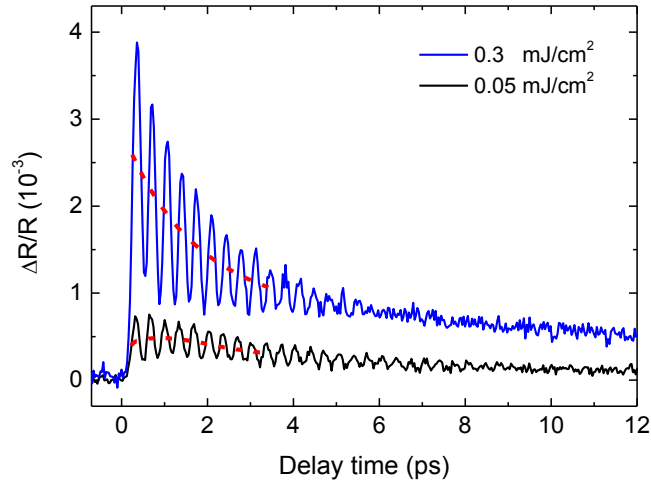

Fig. 5. Decay traces measured using pump fluences of 0.3 and 0.05 mJ/cm$^2$ at 800 nm probe wavelength. Dashed lines indicate the difference of monotonic components at early delay times. The slower rise at low excitation intensity reflects initial electronic dynamics, which limits the efficiency of coherent phonon generation (see the text).

Main features of the photoinduced response in the limit of low excitation intensity are illustrated in Fig. 5. The decay trace measured using pump laser fluence of about 0.05 mJ/cm$^2$ ($N \sim 5\cdot10^{18}$ cm$^{-3}$) demonstrates a delayed rise of the monotonic component. Similar behavior was observed in the experiments with Ti:Saphhire oscillator pulses (see e.g. [2, 3, 44, 45]) and also in THz studies [39]. The different character of initial electronic dynamics as compared to the highly excited state of bismuth is probably caused by low efficiency of electron-electron scattering (including phonon-assisted processes) at moderate densities of photoexcited charge carriers. In this case the filling of energy states near high symmetry points of the Brillouin zone is slower and the signal will appear with a delay.

*4.3. Measurements with variable pump wavelength*

Almost all experiments, in which the delayed response was reported, had been performed using bismuth films with a thickness of ~100 nm or even less [2, 3, 39, 44, 46]. The appearance of decay traces can be then significantly altered, because the relative contribution of such processes as recombination, diffusion, and photoinduced strain relaxation changes considerably with respect to bulk material [47]. In the present work we give a special attention to the diffusion as a possible reason of observed reflectivity kinetics.

The ground state ambipolar diffusion coefficient $D$ for bismuth is 100 cm$^2$/s [29]. However, interactions between the charge carriers in dense electron-hole plasma (for $N \sim 10^{20}$ cm$^{-3}$ and higher) can significantly decrease this value. At the moment, there is a consensus that for a relatively high density of excitation (absorbed laser fluence ~ 1 mJ/cm$^2$ - similar to our measurements) diffusion coefficient reduces to $D \sim 1$ cm$^2$/s [29, 48]. Using ellipsometric data provided by Lenham et al. [49] we estimated the penetration depth $l$ for 400 nm laser radiation to be as large as 30 nm. Then the characteristic time of ambipolar diffusion in laser-excited bismuth is $\tau = l^2/D \sim 10$ ps, which is comparable to the intermediate relaxation time $\tau_2 = 7$ ps. However, we believe that for picosecond electronic dynamics in highly excited bismuth diffusion plays a minor role. This belief is based on the comparison of decay traces measured at a number of excitation wavelengths ranging from 400 nm to 2300 nm.

Indeed, let us consider data presented in Fig. 6 illustrating changes of the photoinduced response that occur upon pump wavelength tuning. Note that the penetration length $l$ increases at wavelengths $\lambda > 1800$ nm or, in other words, the photons with energies less than ~ 0.7 eV are less absorbed inside a bismuth crystal, see Fig. 6(c). The variation of the pump wavelength changes spatial gradient of excited charge carrier distribution, and therefore, the effects caused by diffusion become more or less pronounced depending on the gradient value. However, to obtain a reliable comparison of decay traces one should make sure that the energy of pump radiation absorbed by the probed region of the crystal is the same at every excitation wavelength used (as it was performed in our analysis). From Fig. 6(a) it is clear that the increase of the excited volume results in almost no effect on the character of intermediate $\sigma_2$ relaxation. Since such an increase would slow $\sigma_2$ decay if it was caused by diffusion, it is possible to conclude that the diffusion has minor effects on the ultrafast electronic relaxation in bismuth.

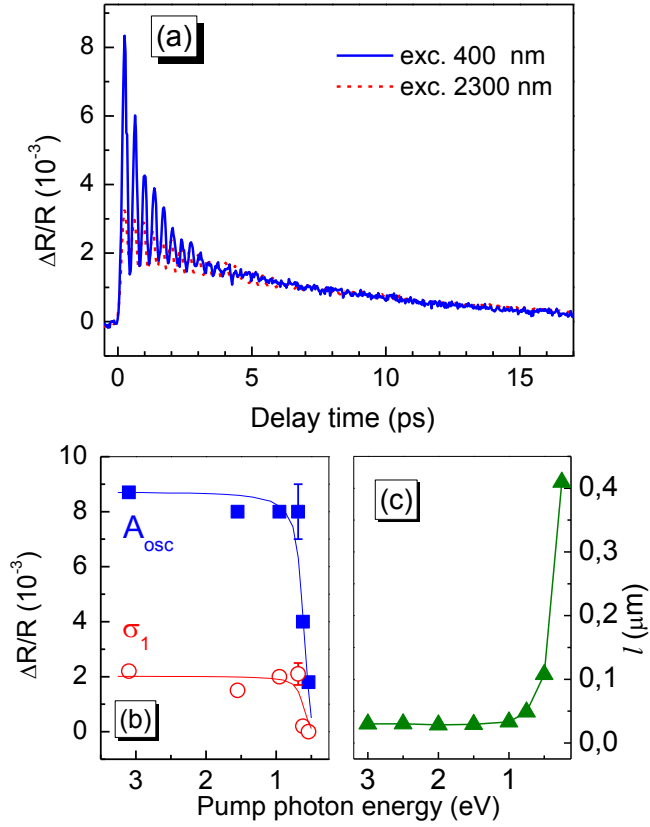

Fig. 6. (a) – Decay traces measured with 400 nm and 2300 nm (dashed line) excitation pulses. (b) – Pump photon energy dependence of the initial amplitude of oscillations ($A_{osc}$) and the amplitude of the 1 ps component ($\sigma_1$). Certain temporal resolution and the duration of pump pulses were taken into account. Solid lines are guides to the eye. (c) – Penetration depth $l$ of bismuth as a function of photon energy calculated using the data provided by [49, 50].

One more important result of measurements with variable pump wavelength is a threshold behavior of several parameters of the photoinduced response. Indeed, one can see that when the pump photon energy is lower than ~ 0.7 eV both the contribution of the fast $\sigma_1$ component and the amplitude of coherent $A_{1g}$ phonons show a considerable decrease. These additional spectral data show once more that nonequilibrium electronic density is segregated into groups occupying different energy states. The observed pump wavelength dependence suggests that charge carriers having slower dynamics reside close to the Fermi level (as in metals). At the same time, the fast $\sigma_1$ component can be linked to the higher-lying electronic states, separated from the valence band by a gap as wide as ~ 0.7 eV.

*4.4. Dynamics of low-energy states*

As follows from our experimental data, $\Delta R/R$ at long time delays remains nonzero, which is due to the long-lived $\sigma_3$ component. Such a behavior is typical for semiconductors, where the electrons excited by an ultrashort laser pulse into the conduction band quickly occupy states near its bottom and characteristic time of the following electron-hole recombination usually lies on the nanosecond scale. Even though bismuth is not a semiconductor, but a semimetal with overlapping conduction and valence bands, electron-hole recombination in it is rather slow, because the overlap is indirect [51]. Therefore, $\sigma_3$ component can be ascribed to nonequilibrium electrons and holes that occupy states in the conduction band near the Fermi level. The time interval $\Delta\tau$ within which these photoinduced charge carriers appear is rather short and is defined by the minor of two quantities: pump pulse duration and characteristic time of thermalization in the individual bands. Regarding temporal resolution achieved in our experiments (100-250 fs) and the high intensity of excitation, it is reasonable to estimate $\Delta\tau$ as ~100 fs. A confirmation of this interpretation is based on our wavelength resolved data and fundamental properties of bismuth.

An interesting experimental fact is that the shape of $\sigma_3(\lambda)$ is similar to the spectral dependence of coherent amplitude $A(\lambda)$: as follows from Fig. 7 both curves have a minimum near 620 nm. This similarity implies a connection between fully symmetric motion of atoms in bismuth and electronic band structure near the Fermi level. Indeed, due to specific crystal structure, derived from the cubic one by small deformations, bismuth can be treated as a Peierls-distorted system [8–11, 19, 27, 31, 40, 52]. As the trigonal axis is concerned, it is possible even to draw an analogy to the one-dimensional case, because the atoms of bismuth form pairs in this direction. The order parameter for this low symmetry state is the internal shift $\delta$ between two cubic sublattices, or what is equivalent – the inverse gap (the energy separation between hole and electron bands). The fully symmetric $A_{1g}$ phonons correspond to out-of-phase oscillations of the two atoms in the unit cell along trigonal axis. These vibrations then represent the Peierls distortion mode and modulate the order parameter of semimetallic phase.

As long as the internal shift is equivalent to the magnitude of indirect band overlap near the Fermi level, its variation leads to the change of free charge carrier density. Since the number of free carriers is among parameters which define the amount of light reflected by the crystal, the internal shift modification $\Delta\delta$ provides the increase or the decrease of reflectivity $\Delta R$. $\Delta R$ is expected to have intra- and interband contribution. The former is due to the change of conductivity, while the latter is caused by the modified population of energy levels involved in optical transitions.

Coherent phonons of $A_{1g}$ symmetry correspond to the periodic variation of $\delta$ and therefore reveal themselves as reflectivity oscillations with wavelength-dependent amplitude $A(\lambda)$. The intraband contribution is most likely to reveal itself as relatively flat $\Delta R$, which is in part confirmed by the fact that the first coherent movement of atoms after pump pulse arrival causes positive $\Delta R$ at every probe wavelength used. A broad minimum of the $A(\lambda)$ curve, clearly seen near 620 nm (2 eV) on Fig. 7, can be ascribed to the effect of interband transitions at the $L$ point of the Brillouin zone (probably it is $L_a(5) - L_s(7)$ transition, the figures in brackets are numbers of bands in the same notation as in [53]).

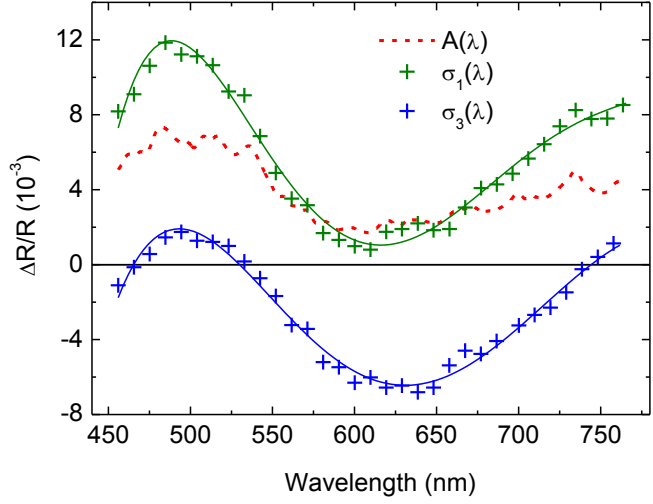

Fig. 7. Spectra of the amplitude of coherent oscillations $A(\lambda)$ and of the fast ($\sigma_1$) and quasiconstant ($\sigma_3$) components of the photoinduced response. Solid lines are guides to the eye.

Apart from internal shift variation, promotion of additional electrons to the *L* point (or holes to *T* point) by light quanta also changes the number of free charge carriers. Under femtosecond laser excitation it is possible because excited electrons lose their energy and accumulate partially near the Fermi level. The contribution of these charge carriers to the photoinduced response is represented by the nanosecond component $\sigma_3$, whose spectrum should then resemble the wavelength-dependent $A_{1g}$ amplitude. Exactly this is observed in our measurements, where $\sigma_3(\lambda)$ and $A(\lambda)$ are similar.

The state of photoexcited charge carriers accumulated near the Fermi level might be in principle characterized by a certain temperature, which is higher than the lattice temperature. In this context we believe that a metal-like transfer of energy from the excited electrons to the lattice due to electron-phonon interaction reveals itself in bismuth as the $\sigma_2$ component of the photoinduced response. As already mentioned, this component exhibits a decrease of reflectivity, which is uniform across all visible spectrum ($\sigma_2$ is wavelength-independent) and is characterized by 7 ps decay time. It is well known that the reflectivity of bismuth gradually decreases for a higher temperature and the decrease is almost spectrally uniform [54]. Therefore, we can ascribe $\sigma_2$ to the heating of the crystal via interaction of lattice with a part of excited electrons. Using data provided by [37], it is possible to estimate the temperature increase. Since $d(\Delta R/R)/dT$ is $8.5*10^{-5}$ $K^{-1}$, then dividing the amplitude of the $\sigma_2$ component (which is 0.007) by this value we estimate that $\Delta T$ is ~ 80 K. Therefore, after about 30 ps, when the heating process is over, the temperature is ~ 380 K, which is significantly lower than the melting temperature of bismuth (490 K). This estimate shows that for the pump energy densities used in our experiments ($\Phi$ ~ 1 mJ/cm$^2$) bismuth remains crystalline during relaxation, and therefore the model of an overheated lattice proposed for the same intensities of excitation is inapplicable [20].

*4.5. Electronic dynamics at T point of the Brillouin zone*

In the previous sections we have discussed cooling and nanosecond recombination of charge carriers. However, the monotonic part of the photoinduced response also contains $\sigma_l$ relaxation component with a considerably higher decay rate. For the interpretation of $\sigma_l$ in bismuth it is useful again to refer to coherent lattice dynamics. A key result here is that the lifetime of $\sigma_l$ component ($\tau_l \approx 1$ ps) coincides within experimental errors with the characteristic time of frequency shift relaxation. While $\sigma_l$ is controlled by the dynamics of charge carriers, the instantaneous frequency of $A_{1g}$ phonon mode reflects the current state of the crystal lattice, and such a coincidence can be an evidence of a strong coupling between the electronic and vibrational degrees of freedom.

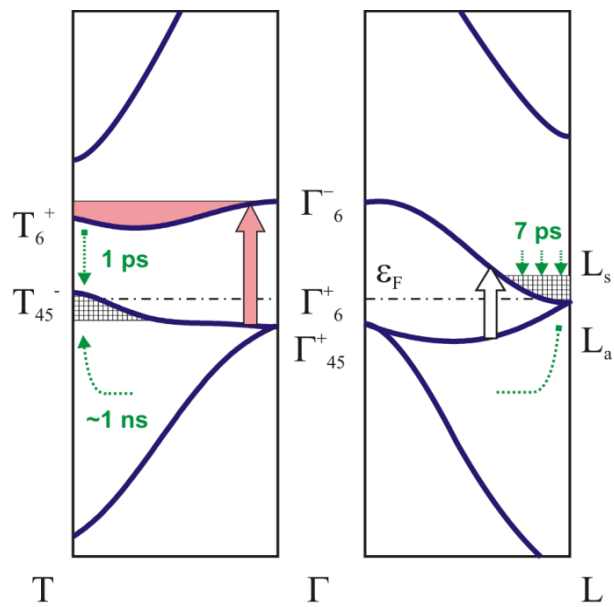

Fig. 8. Part of the band structure of bismuth adapted from [53]. Filled arrow shows resonant excitation of electrons leading to the appearance of fast 1 ps component and effective generation of coherent $A_{1g}$ phonons (see the text). Empty arrow stands for "below-threshold" transitions which cause the accumulation of charge carriers near the Fermi level. Dashed regions corresponds to excited electrons (*L* point) and holes (*T* point) having nanosecond lifetime. Dotted lines are transitions leading to the relaxation of photoexcited species (recombination and cooling).

The short-lived component $\sigma_l$ represents a portion of excited charge carriers located near the particular point of the Brillouin zone. For $\sigma_l$ this location can be found relying upon the data obtained in the experiments with tunable excitation photon energy. Indeed, the amplitude of $\sigma_l$ experiences a sharp drop when photon energy is in the near infrared range (see Fig. 6(b)). To be more specific, we characterize this behavior of $\sigma_l$ by a threshold value of 0.7 eV. To relate $\sigma_l$ to the electronic properties of bismuth we referred to the band structure of bismuth suggested by Golin [53].

The value of 0.7 eV is close to the energies of two optical transitions $\Gamma_6^+(5) - \Gamma_6^-(6)$ and $\Gamma_{45}^+(4) - \Gamma_6^-(6)$ which are 0.821 eV and 0.825 eV respectively (according to [53]). However, since the piezoreflectance measurements [50] found the corresponding energies to be 0.69 eV and 0.81 eV, we associate the component $\sigma_1$ with the excited electrons that occupy states in the band containing $\Gamma_6^-$ point, see Fig. 8. Therefore, if the photon energy is lower than the gap between $\Gamma_6^+$ and $\Gamma_6^-$ points, the electrons are not effectively promoted to this band and $\sigma_1$ remains small. If the pump photon energy is 0.7 eV or higher, the electrons are able to occupy these states directly, or via the interactions with other excited charge carriers during the energy redistribution in the electronic system.

Here we would like to emphasize similar excitation wavelength dependences of $\sigma_1$, the coherent $A_{1g}$ amplitude and the magnitude of frequency shift $\Delta\phi$ (see Fig. 6 and Fig. 9). This experimental fact allows to link the softening of $A_{1g}$ mode to the population of the band containing $\Gamma_6^-$ point and resembles general features of coherent phonon generation in semiconducting crystals. Indeed, semiconductors are characterized by the valence and conduction bands formed by electronic states of different configurations, which can be described as bonding and anti-bonding. As a result, the above band gap excitation by ultrafast laser pulses leads to the weakening of interatomic forces and causes coherent lattice oscillations. The larger is the number of electrons promoted to the conduction band – the weaker is the interatomic binding and the lower is the frequency of coherent phonons.

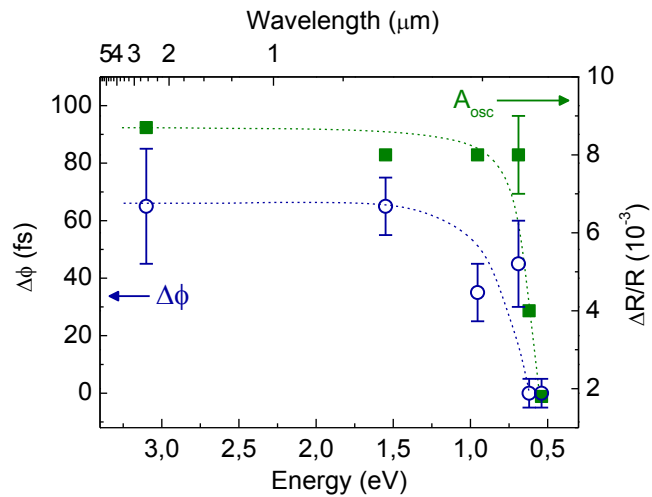

Fig. 9. Pump photon energy dependence of the initial amplitude $A_{osc}$ and of the relative phase shift $\Delta\phi$ of coherent oscillations. Dashed lines are guides to the eye.

For bismuth the situation is, however, more specific. The gap between "valence" and "conducting" bands along the $\Gamma$-$T$ section is formed upon folding of electronic dispersion curves, which in turn is caused by the pairing of atoms and the halving of the unit cell along $\Gamma$-$L$-$T$ direction. Therefore, in analogy with quasi 1D materials, the softening of the LO phonon mode ($A_{1g}$ mode in this case) is due to its strong interaction with the electronic states at the edge of the Brillouin zone (for bismuth – near $T$ point). Such "singularity" of electron-phonon interaction stems from the electronic instability inherent for hypothetic cubic structure of bismuth in the absence of Peierls distortion. Therefore, our analysis will be more consistent if we locate the excited $\sigma_1$ electrons near the $T_6^+$ point of the Brillouin zone. Ultrafast accumulation of electrons in this region with high density of states can be then regarded as the source of coherent atomic motion and the origin of $A_{1g}$ mode softening in bismuth. According to this model the strongest coupling is observed between phonon states near the center ($\Gamma$ point) and electronic states at the edge of the Brillouin zone ($T$ point). Therefore, the frequency softening is related only to the small part of the $A_{1g}$ phonon branch (this question has been debated in [17, 55]).

Taking in account the above reasoning, one can conclude that coherent phonons represent damped harmonic oscillations of atoms near their instantaneous shifted equilibrium position $q$ which returns to its unperturbed value $q_0$ exponentially with ~1 ps characteristic time ($q(t) \approx q_0 - q(0)e^{-t}$). In an ideal situation, when the photoinduced displacement is much faster than the inverse period of $A_{1g}$ mode, the oscillations are cosine with the amplitude $A = q_0 - q(0)$. In some respect this suggestion is consistent with the results of optical pump-probe measurements, which show the similarity of $A(\lambda)$ and $\sigma_1(\lambda)$ spectra. In practice, the efficiency of coherent phonon generation is limited either by a finite duration of a pump pulse, or by characteristic time of energy redistribution in the electronic system. The second factor is clearly evident at low pump intensities, when electron-electron scattering is less efficient due to a moderate density of photoexcited charge carriers, see Fig. 5.

Finally, we make a few remarks concerning the nature of picosecond $\sigma_1$ relaxation. One may try to interpret this process by introducing independent Fermi-Dirac distributions for excited electrons at the $T_6^+$ and holes at the $T_{45}^-$ point at the end of a pump pulse. The decay of $\sigma_1$ component then corresponds to thermodynamic equilibration of these species through intervalley and interband scattering. However, such an approach has several disadvantages. First of all, at high densities of photoexcited charge carriers electron-electron scattering rates in bismuth substantially exceed the 1 ps$^{-1}$ value. Secondly, the electrons at the $L$ point can have their own temperature and chemical potential due to specific form of Fermi surface of bismuth. Therefore, the concept of individual thermal distributions turns out to be rather complex, because one should deal with at least three groups of photoexcited charge carriers. Additionally, it is not clear how the change of any thermodynamic quantity related to electronic system can lead to the softening of a particular phonon mode [56].

Thus, it is reasonable to treat $\sigma_1$ decay just as the relaxation of electronic population. Since indirect $L$-$T$ recombination is characterized by nanosecond time, we suggest the direct recombination of electrons at the $T_6^+$ point with holes at the $T_{45}^-$ point. Transitions between these states conserve momentum, while additional quasiparticles are needed to carry away the excess energy. Here we can treat these species as optical phonons. At first glance it may seem that the process has a low efficiency, because the energy of each electron-hole pair is much higher compared to the energy of optical phonon. However, one should notice that electronic states near the $T$ point are strongly coupled to $A_{1g}$ phonons. Therefore, Peierls distorted nature of a bismuth crystal can greatly enhance the probability of phonon-assisted recombination.

The energy of an electron-hole pair at the *T* point (~ 0.5 eV [53]) is then corresponds to the energy of about 40 $A_{1g}$ phonons (0.012 meV), which are emitted incoherently.

## 5. Summary

In conclusion, we have measured ultrafast photoinduced response of crystalline bismuth using a wide range of wavelengths in the visible and near infrared both for excitation and probing. In contrast to simple models, which characterize the excited state of bismuth by the total number of excited electrons, we observed a diverse electronic dynamics spanning three orders of magnitude in the time domain – from pico- to nanoseconds. The analysis of time-resolved reflectivity signal in all visible range revealed several groups of excited charge carriers with different coupling to the lattice. The shortest electronic decay rate obtained is 1 ps$^{-1}$ and coincides with the relaxation rate of the $A_{1g}$ frequency shift. The observed simultaneous decrease of coherent amplitude and the contribution of the fastest component at pump photon energies below 0.7 eV allowed us to single out a portion of excited electrons strongly coupled to the fully symmetric lattice motion. Using an analogy to Peierls-distorted systems, we ascribed the generation of coherent $A_{1g}$ phonons to ultrafast population of electronic states near the *T* point of the Brillouin zone.

The incoherent dynamics in bismuth include lattice heating and a relatively long indirect electron-hole recombination near the Fermi level. While the former results in a spectrally uniform signal, the low energy charge carriers induce reflectivity changes that resemble the spectra caused by atomic vibrations of $A_{1g}$ symmetry. This similarity is ascribed to the modulation of the band separation by $A_{1g}$ phonons and is consistent with the concept of a Peierls-distorted crystal.

## Acknowledgments

Authors acknowledge the support of Russian Foundation of Basic Research (project #12-02-00898-a), and thank V. O. Kompanets for the help with experiments, and S. I. Kudryashov for stimulating discussions.

## References

[1]   Cheng T K, Brorson S D, Kazeroonian A S, Moodera J S, Dresselhaus G, Dresselhaus M S and Ippen E P 1990 *Appl. Phys. Lett.* **57** 1004

[2]   Hase M, Mizoguchi K, Harima H and Nakashima S 1996 *Appl. Phys. Lett.* **69** 2474

[3]   Hase M, Mizoguchi K, Harima H, Nakashima S-ichi and Sakai K 1998 *Phys. Rev.* B **58** 5448

[4]   DeCamp M, Reis D, Bucksbaum P and Merlin R 2001 *Phys. Rev.* B **64** 092301

[5]   Hase M, Kitajima M, Nakashima S-ichi and Mizoguchi K 2002 *Phys. Rev. Lett.* **88** 067401

[6]   Sokolowski-Tinten K, Blome C and Blums J 2003 *Nature* **422** 287

[7] Sokolowski-Tinten K and Linde D V D 2004 *J. Phys.: Cond. Matt.* **16** R1517

[8] Sciaini G, Harb M, Kruglik S G, Payer T, Hebeisen C T, zu Heringdorf F-J M, Yamaguchi M, Horn-von Hoegen M, Ernstorfer R and Miller R J D 2009 *Nature* **458** 56

[9] Fal'kovski L A 1968 *Sov. Phys. Usp.* **11** 1

[10] Édel'man V S 1977 *Sov. Phys. Usp.* **20** 819

[11] Shick A, Ketterson J, Novikov D and Freeman A 1999 *Phys. Rev.* B **60** 15484

[12] Misochko O, Hase M, Ishioka K and Kitajima M 2004 *Phys. Rev. Lett.* **92** 197401

[13] Diakhate M S, Zijlstra E S and Garcia M E 2009 *Appl. Phys.* A **96** 5

[14] Misochko O V, Hase M, Ishioka K and Kitajima M 2004 *Phys. Lett.* A **321** 381

[15] Misochko O and Lebedev M *Chin. J. Phys.* **49** 141

[16] Johnson S, Beaud P, Vorobeva E, Milne C, Murray É, Fahy S and Ingold G 2009 *Phys. Rev. Lett.* **102** 175503

[17] Zijlstra E S, Díaz-Sánchez L E and Garcia M E 2010 *Phys. Rev. Lett.* **104** 135701

[18] Misochko O, Sakai K and Nakashima S 2000 *Phys. Rev.* B **61** 11225

[19] Zeiger H, Vidal J, Cheng T, Ippen E, Dresselhaus G and Dresselhaus M 1992 *Phys. Rev.* B **45** 768

[20] Boschetto D, Gamaly E G, Rode A V, Glijer D, Garl T, Albert O, Rousse A and Etchepare J 2008 *Phys. Rev. Lett.* **100** 027404

[21] Giret Y 2011 *Phys. Rev. Lett.* **106** 155503

[22] Li L, Checkelsky J G, Hor Y S, Uher C, Hebard A F, Cava R J and Ong N P 2008 *Science* **321** 547

[23] Behnia K, Balicas L and Kopelevich Y 2007 *Science* **317** 1729

[24] Tediosi R, Armitage N, Giannini E and van der Marel D 2007 *Phys. Rev. Lett.* **99** 016406

[25] Chudzinski P and Giamarchi T 2011 *Phys. Rev.* B **84** 125105

[26] Zhu Z, Collaudin A, Fauqué B, Kang W and Behnia K 2011 *Nature Physics* **8** 89

[27] Fritz D M et al. 2007 *Science* **315** 633

[28] Esmail A R and Elsayed-Ali H E 2011 *Appl. Phys. Lett.* **99** 161905

[29] Johnson S, Beaud P, Milne C, Krasniqi F, Zijlstra E, Garcia M, Kaiser M, Grolimund D, Abela R and Ingold G 2008 *Phys. Rev. Lett.* **100** 155501

[30] Moriena G, Hada M, Sciaini G, Matsuo J and Dwayne Miller R J 2012 *J. Appl. Phys.* **111** 043504

[31] Papalazarou E, Faure J, Mauchain J, Marsi M, Taleb-Ibrahimi A, Reshetnyak I, van Roekeghem A, Timrov I, Vast N, Arnaud B and Perfetti L 2011 *arXiv: 1112.3949*

[32] Fahy S and Reis D 2004 *Phys. Rev. Lett.* **93** 109701

[33] Murray É, Fritz D, Wahlstrand J, Fahy S and Reis D 2005 *Phys. Rev.* B **72** 060301

[34] Mel'nikov A A, Misochko O V and Chekalin S V 2009 *JETP Letters* **89** 129

[35] Garl T, Gamaly E, Boschetto D, Rode A, Luther-Davies B and Rousse A 2008 *Phys. Rev.* B **78** 134302

[36] Allen P B 1987 *Phys. Rev. Lett.* **59** 1460

[37] Wu A Q and Xu X 2007 *Appl. Phys. Lett.* **90** 251111

[38] Zhou P, Rajkovi I, Ligges M and Payer T 2009 *Ultrafast Phenomena XVI* ed Corkum P, De Silvestri S, Nelson K A, Riedle E, Schoenlein R W (Berlin: Springer)

[39] Timrov I, Kampfrath T, Faure J, Vast N, Ast C, Frischkorn C, Wolf M, Gava P and Perfetti L 2012 *Phys. Rev.* B **85** 155139

[40] Murray É, Fahy S, Prendergast D, Ogitsu T, Fritz D and Reis D 2007 *Phys. Rev.* B **75** 184301

[41] Zijlstra E, Tatarinova L and Garcia M 2006 *Phys. Rev.* B **74** 220301

[42] Johnson S L, Beaud P, Vorobeva E, Milne C J, Murray É D, Fahy S and Ingold G 2010 *Phys. Rev. Lett.* **104** 220301

[43] Misochko O V, Ishioka K, Hase M and Kitajima M 2006 *J. Phys.: Cond. Matt.* **18** 10571

[44] Hase M, Ishioka K, Kitajima M and Ushida K 2000 *Appl. Phys. Lett.* **76** 1258

[45] Ishioka K, Kitajima M and Misochko O V 2006 *J. Appl. Phys.* **100** 093501

[46] Wu A Q and Xu X 2007 *Appl. Surf. Sci.* **253** 6301

[47] Sheu Y-M 2010 *Ultrafast Dynamics of Photoexcited Bismuth Films* (The University of Michigan)

[48] Hase M, Kitajima M, Nakashima S-ichi and Mizoguchi K 2004 *Phys. Rev. Lett.* **93** 117401

[49] Lenham A and Treherne D 1965 *J. Opt. Soc. Am.* **231** 1964

[50] Wang P and Jain A 1970 *Phys. Rev.* B **2** 2978

[51] Lopez A 1968 *Phys. Rev.* **175** 823

[52] Misochko O V, Lu R, Hase M and Kitajima M 2007 *J. Exp. Theor. Phys.* **104** 245

[53] Golin S 1968 *Phys. Rev.* **166** 643

[54] Cardona M and Greenaway D 1964 *Phys. Rev.* **133** A1685

[55] Murray É, Fahy S, Prendergast D, Ogitsu T, Fritz D and Reis D 2007 *Phys. Rev.* B **75** 184301

[56] Tangney P and Fahy S 2002 *Phys. Rev.* B **65** 054302